\documentclass[twocolumn,times]{emulateapj}

\usepackage{color,ulem}
\usepackage{multirow}
\usepackage{amsmath}
\usepackage{mathdots}
\usepackage{bm}
\usepackage[hidelinks=true]{hyperref}

\newcommand{\dtau}{\delta\tau}

\shorttitle{Computing Helioseismic Sensitivity Kernels}
\shortauthors{Hartlep \& Zhao}

\begin{document}

\title{Computing Helioseismic Sensitivity Kernels \\ for the Sun's Large-Scale Internal Flows using Global-Scale Wave-Propagation Simulations}

\author{Thomas Hartlep}
\affiliation{Bay Area Environmental Research Institute, Moffett Field, CA 94035, USA}
\affiliation{NASA Ames Research Center, Moffett Field, CA 94035, USA}

\author{Junwei Zhao}
\affiliation{W.W. Hansen Experimental Physics Laboratory, Stanford University, Stanford, CA 94305-4085, USA}

\begin{abstract}

Helioseismic waves observable at the solar surface can be used to probe the properties of the Sun's interior.
By measuring helioseismic travel times between different location on the surface, flows and other interior properties can be inferred using so-called sensitivity kernels which relate the amount of travel-time shift with variations in interior proporties.
In particular, sensitivity kernels for flows have been developed in the past, using either ray or Born approximation, and have been used to infer solar interior flows such as the meridional circulation which is of particular interest for understanding the structure and dynamics of the Sun.
Here we introduce a new method for deriving three-dimensional sensitivity kernels for large-scale horizontal flows in the solar interior.
We perform global-Sun wave-propagation simulations through 784 small flow perturbations placed individually in the interior of a simulated Sun, and measure the shifts in helioseismic travel times caused by these perturbations.
Each measurement corresponds to a linear equation connecting the flow perturbation velocities and the sensitivity kernels.
By solving the resulting large set of coupled linear equations, we derive three-dimensional sensitivity kernels for horizontal flows which have a longitudinal component (parallel to the wave's travel direction) and a transverse component (perpendicular to the wave's travel direction).
The kernels exhibit a ``banana'' shape, similar to kernels derived using Born approximation methods, and show that transverse components are not negligible in inversions for interior flows.

\end{abstract}


\section{Introduction} \label{sec:introduction}
Time--distance helioseismology \citep{Duvalletal1993} measures the travel times of acoustic waves traveling between different locations on the solar surface.
Since helioseismic waves travel from one surface location to another through the solar interior, their travel times are affected by interior properties of the Sun such as sound speed, mass flows and magnetic fields.
In principle, these properties can be inferred by inverting measured travel times \citep{Kosovichev1996}.
In particular, by measuring variations or shifts in helioseismic travel times relative to some reference model, one can measure perturbations in solar interior properties.
Assuming that these perturbations are small, a linear convolution equation links travel-time shifts, $\dtau$,
measured between two surface locations $(\theta_{\cal S}, \phi_{\cal S})$ and $(\theta_{\cal R}, \phi_{\cal R})$ with perturbations of some interior property, $\delta_{\cal Q}$:
\begin{equation}
\begin{split}
   & \dtau(\theta_{\cal S}, \phi_{\cal S}; \theta_{\cal R}, \phi_{\cal R}) = \\
   & \int \left[ \hat{D}(\theta_{\cal S}, \phi_{\cal S}; \theta_{\cal R}, \phi_{\cal R})K^{(d)}_{\cal Q} \right](\bm{r})
    \delta_{\cal Q}(\bm{r}) dV_{\bm{r}},
\label{eqn:generickernel}
\end{split}
\end{equation}
where $\bm{r}=(r, \theta, \phi)$ and $dV_{\bm{r}}$ are the spherical coordinates (radius, co-latitude, and longitude) inside the Sun and the corresponding volume element, $K^{(d)}_{\cal Q}$ is the travel-time sensitivity kernel for property $\cal Q$, and $\hat{D}$ is an operator that rotates the kernel such that its source and receiver locations (see section~\ref{sec:kernels:discrete}) are at the two points between which travel-time shifts are measured.
Written in this way, the kernel $K^{(d)}_{\cal Q}$ only depends on the distance $d$ between the two points.
In general, measured travel-time shifts represent the cumulative effect of different interior properties (e.g., sound speed variations, flows, magnetic fields), each of which have their own sensitivity kernel.
For accurate inferrence of the Sun's internal structures and dynamics, accurate models of these kernels are of great importance.
In this paper, we only consider the effect of mass flows but our method can in principle be applied to other internal properties as well.

The simplest sensitivity kernels for flows can be derived using a ray approximation which assumes that the Sun's acoustic waves have infinitely small wavelengths.
Many authors have employed ray-approximation kernels to infer subsurface properties of sunspots \citep[e.g.,][]{Kosovichev1996, Zhaoetal2001, Couvidatetal2004} and the Sun's large-scale flows \citep[e.g.,][]{ZhaoKosovichev2004}.
Despite their simplicity and oversimplified assumption, ray-approximation methods have provides tremendous insight into the structure and dynamics of the solar interior.
Meanwhile, Born-approximation sensitivity kernels have also been derived, taking into account the finite wavelength and scattering of acoustic waves inside the Sun \citep{GizonBirch2002, Birchetal2004, BirchFelder2004}.
Subsurface flows have been inverted using such kernels \citep[e.g.,][]{Svandaetal2011}.
However, early works derived Born-approximation sensitivity kernels in Cartesian coordinates, and have therefore focused only on the Sun's shallow interior.

More recently, with the discovery of a systematic center-to-limb effect in the helioseismic measurements \citep{Zhaoetal2012, ChenZhao2018}, efforts on detecting deep equatorward meridional flow inside the Sun and inferring the Sun's full internal meridional-circulation profile have gained momentum.
\citet{Zhaoetal2013} first reported a detection of an equatorward flow residing in the middle of the convection zone between about 0.83~R$_\sun$ and 0.91~R$_\sun$ in each hemisphere.
Later analyses by various authors, following the same or similar analysis procedures, all reported detection of equatorward flows even though their inferred meridional-circulation profiles disagreed quantitatively \citep{Kholikovetal2014, Jackiewiczetal2015, RajaguruAntia2015, ChenZhao2017, LinChou2018, Liangetal2018}.
All these authors employed the ray-approximation sensitivity kernels for flows, which are much easier to derive in a global-scale spherical coordinate system.
Realizing the limits of ray-approximation kernels, some authors have started to develop wave-based sensitivity kernels.
For example, \citet{Boeningetal2016} extended a local-scale Born-approximation method to the global scale and derived sensitivity kernels for flows in spherical coordinates; from a different perspective, \citet{Gizonetal2017} developed a more computationally efficient method that calculated the sensitivity kernels under the first Born approximation with better numerical precision; and \citet{Mandaletal2017} calculated finite-frequency sensitivity kernels through building Green's functions.
These wave-based sensitivity kernels have also been used in inversions to derive the Sun's meridional circulation profile \citep{Boeningetal2017, Mandaletal2018}, which gave results that may be more reliable but still rather similar to those obtained from ray-approximation methods.
Complementary to time-distance helioseismology, there are also global helioseismology techniques that have been developed for measuring deep meridional flows, such as
studies based on meridional-flow perturbed eigenfunctions of the solar $p$-modes \citep{Schadetal2013}.

The aforementioned sensitivity kernels are based on either ray approximation or Born approximation.
In this article, we introduce a method for calculating sensitivity kernels for the Sun's large-scale internal flows by performing a set of global wave-field simulations with small flow perturbations placed at different locations and depths inside a solar model.
Although the idea of deriving sensitivity kernels through measuring travel-time shifts by localized perturbations was explored previously \citep[e.g.,][]{Duvalletal2006, Hanasogeetal2007}, this is the first time that such a technique has been worked through to infer actual three-dimensional spherical kernels.
Our method complements the existing Born-approximation methods using a completely different approach.
It has the potential to improve the accuracy of flow inferences, or at the very least improve our confidence in the validity and reliability of current Born-approximation kernels.

This article is organized as follows: Section~\ref{sec:methodology} describes the methodology of our approach, Section~\ref{sec:results} presents select kernels obtained with the new approach; checks for self-consistency of the solutions are presented in Section~\ref{sec:validation}; a discussion and future work follow in Section~\ref{sec:discussion}.

\section{Methodology} \label{sec:methodology}
The flow sensitivity kernel at some location $(r, \theta, \phi)$ in the solar interior can be interpreted as the amount of travel-time (or phase) shift observable on the surface between waves traveling in opposite directions caused by an infinitesimally small flow perturbation at that location, normalized by the perturbation volume and its flow speed.
This is because the convolution equation (Equation~\ref{eqn:generickernel}) contracts to $\dtau \sim K_{\cal P}$ for perturbations that are spatial delta-functions.
By simulating wave propagation through infinitesimally small perturbations at many different locations and measuring their affect on travel times, the entire kernel could be mapped out directly.
In practice, of course, we cannot simulate infinitesimally small perturbations, simply for numerical reasons as well as the fact that their affect on travel times would not be measurable.
The closest we can get to this scenario is by using flow perturbations that are small relative to the finite wavelength of helioseismic waves, and with speeds that are small compared to the local sound speed to maintain linearity.
For each perturbation and each measurement location, we then have an equation the like of Equation~\ref{eqn:generickernel}.
The resulting (coupled) set of (linear) equations can then be solved numerically for the sensitivity kernel.

In the following, we describe the helioseismic wave-propagation simulations (Section~\ref{sec:simulations}), the procedure for measuring wave travel-times from the simulated wave-fields (Section~\ref{sec:measurements}), and the method for deriving sensitivity kernels from these measurements (Section~\ref{sec:kernels}).

\subsection{Wave-Propagation Simulations} \label{sec:simulations}

\begin{figure*}[t]
   \centering
   \includegraphics[width=0.65\linewidth]{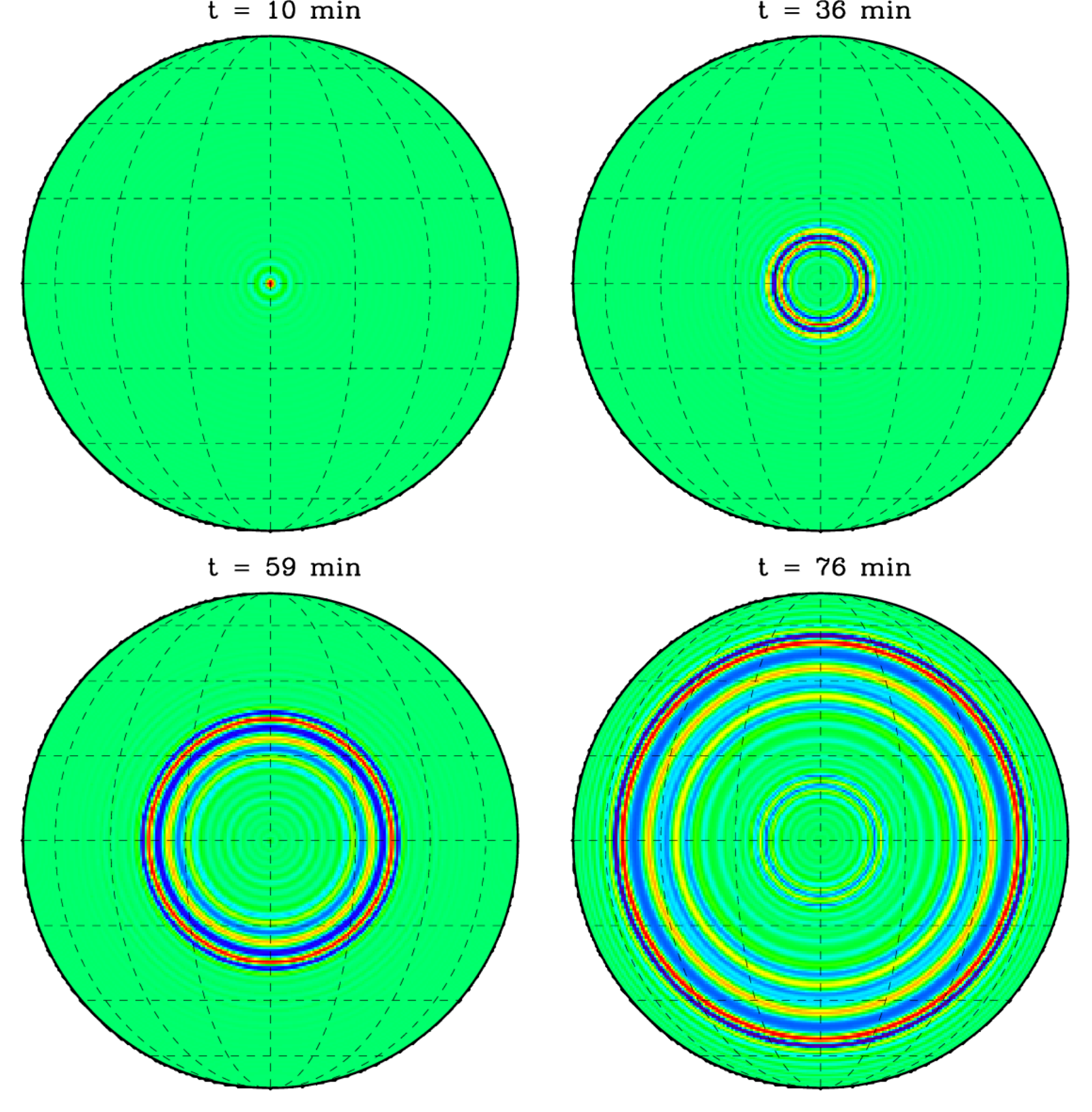}
   \caption{Snapshots of a simulated single-source wave at select moments in time as indicated in the panels.
   Shown is the radial velocity at 300\,km above the photosphere, with green indicating zero velocity, and red and blue indicating positive and negative velocities, respectively.
   \label{fig:snapshots}}
\end{figure*}

\subsubsection{Simulation Code} \label{sec:code}

Central to our method for deriving sensitivity kernels are numerical simulations of the helioseismic wave-field in the presence of localized flow perturbations in the solar interior.
The calculation are carried out by a numerical code we call ``a{\it SHB\/}ury'' which uses a {\it S\/}herical-{\it H\/}armonic--{\it B\/}-spline discretization method for solving the propagation of helioseismic waves in the full 3D solar interior \citep{HartlepMansour2004, HartlepMansour2005}.
The version of the code used here is essentially identical to the one employed by \citet{Hartlepetal2013}.
The code has been used over many years to study wave propagation in the solar interior in the presence of both sound-speed perturbations as well as flows, and was used to help improve and validate various helioseismic measurement and inversion techniques.
This includes the far-side imaging technique for detecting solar active regions on the side facing away from Earth using observations from the Earth-facing side \citep{Zhao2007,Hartlepetal2008,Ilonidisetal2009}; measuring the solar meridional flow in the deep interior \citep[e.g.,][]{Hartlepetal2013, Rothetal2016}; detecting sound-speed variation using time-distance helioseismoloy \citep{Parchevskyetal2014} and the acoustic holography technique \citep{Braun2014,DiazAlfaroetal2016}; and studies related to magnetic flux emergence \citep[e.g.,][]{Hartlepetal2011,Ilonidisetal2013}.

The code models solar acoustic oscillations in a spherical domain using the Euler equations linearized around a stationary background state, which is characterized by a background density, sound speed and acceleration due to gravity, all taken from a standard solar model.
The code also takes into account the effect of mass flows on the propagation of waves.
For the governing equations and more details on the code, we refer to \citet{Hartlepetal2013}.

A significant difference between how the code is used in the currect study and previous works is the excitation of acoustic waves.
In the Sun, turbulent flows near the photosphere act as sources for acoustic perturbations driving waves in a stochastic fashion.
However, since this driving mechanism is a nonlinear process, it is lost by linearizing the governing equations.
Therefore, for most previous studies, a stochastic term was added in the equations to mimic this excitation.
This is not used here, however.
Instead, each simulation excites and follows individual waves as described in the next section.

\subsubsection{Simulation Setup} \label{sec:setup}

In order to measure helioseismic travel times accurately, one usually needs many hours of data, for either actual solar observations or numerical simulations that are stochastically driven to mimic solar-like wave excitation.
Such numerical simulations are feasible when only one or few sets of them are needed, as for instance in \citet{Hartlepetal2008,Hartlepetal2013}.
However, for deriving sensitivity kernels, we need to probe many perturbations at many different depths and locations while keeping them spatially separated to avoid travel times to be shifted by multiple perturbations simultaneously.
This requires many individual simulations, with only few perturbations in each, which quickly becomes infeasible.

The alternative we have chosen here is to simulate not a stoachstic Sun-like wave-field but excite individual waves and measure their travel times from their source location to other points on the surface.
There is no stochastic noise in this case, and since helioseismic waves travel halfway around the Sun in less 150\,min or so, simulations are fairly short and we can place two waves, and two different perturbations, on opposite sides of the Sun to minimize ``cross-talk'' between them.
Specifically, we have chosen to excite two waves at the solar equator 180\degr~apart from each other by initializing the simulation with small, Gaussian-shaped density perturbations.
These perturbations are centered at 200\,km below the photosphere, near the depth at which waves are excited in the Sun, and have a full width at half maximum (FWHM) of 235\,km in depth and $1\fdg66$ horizontally.
Such initial density perturbations launch single waves traveling away from their source locations.
Snapshots of such a wave from one simulation are shown in Figure~\ref{fig:snapshots}.

\begin{figure}[t]
   \centering
   \includegraphics[viewport=18mm 5mm 179mm 100mm,clip,width=\linewidth,height=0.25\textheight,keepaspectratio]{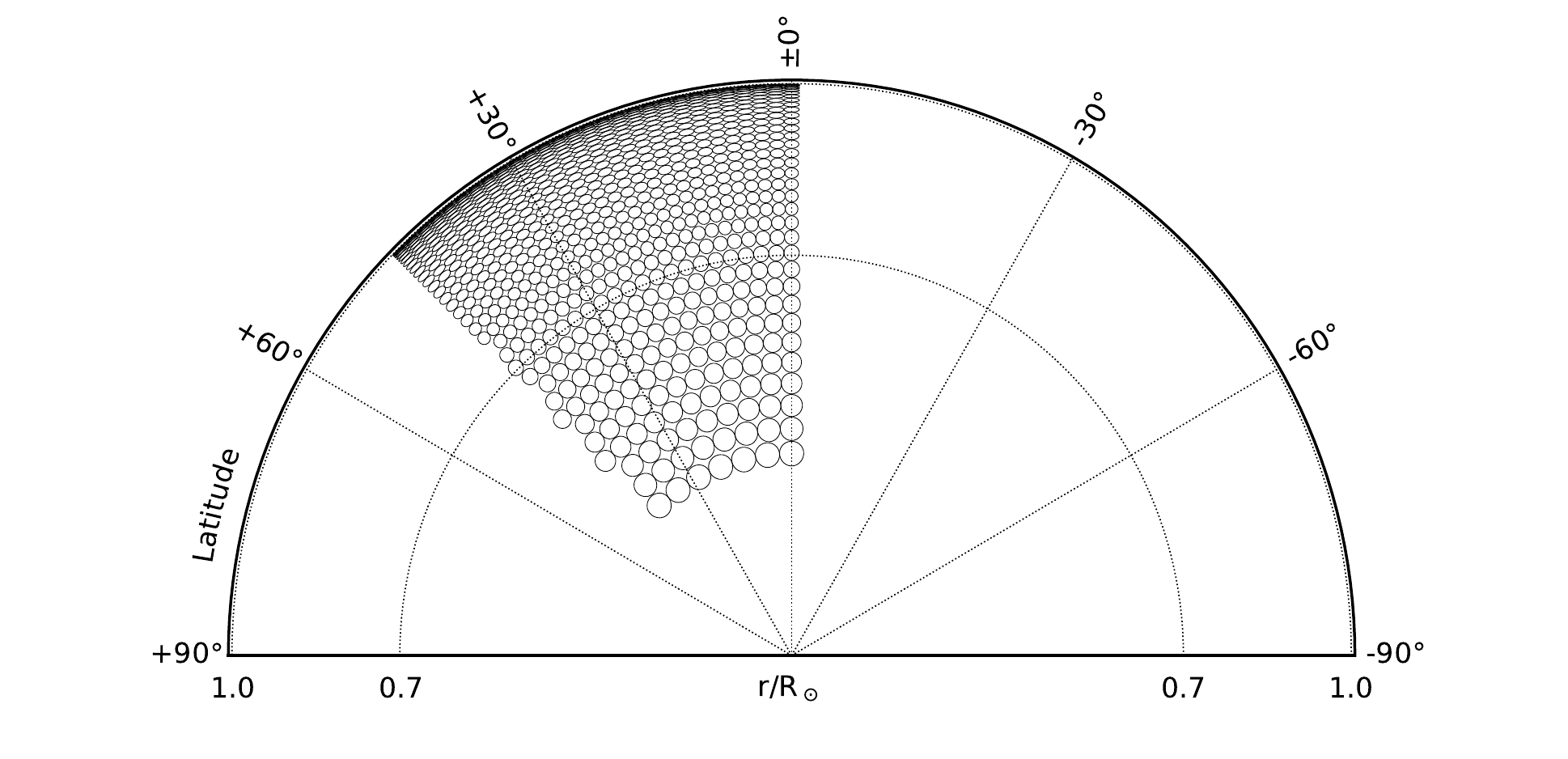}
   \caption{Cross-sectional view of the different locations (depth and latitudinal offset from the wave source) of the 784 Gaussian-shaped flow perturbations used in the wave-propagation simulations.
   In each simulation run, two such perturbations are placed on opposite sides of the simulated Sun.
   The plot shows a single contour for each perturbation at 50\% of their maximum flow amplitude.
   \label{fig:perturbations}}
\end{figure}

\begin{figure}[t]
   \centering
   \includegraphics[viewport=8mm 5mm 182mm 87mm,clip,width=\linewidth,height=0.25\textheight,keepaspectratio]{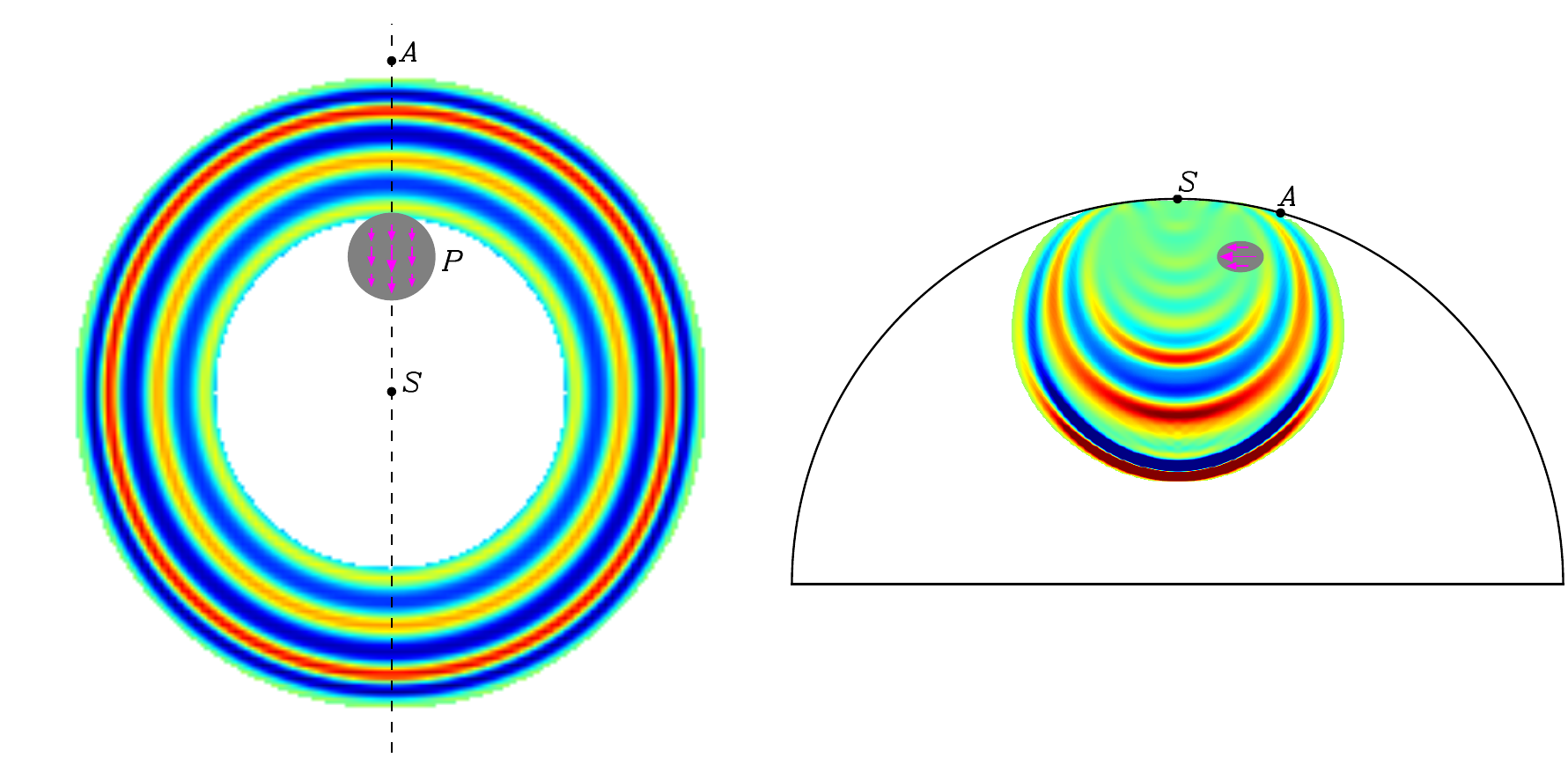}
   \caption{An example of top view (left) and cross-sectional view (right) of the simulated wave-field, in blue and red, together with an internal flow-perturbation volume ${\cal P}$, represented by the grey circle (left) or ellipse (right) with magenta arrows indicating the flow direction.
   The dashed line is the central axis crossing the wave source ${\cal S}$ and in parallel with the flow direction, and the right panel is the cross-sectional view through the axis ${\cal S}$--${\cal A}$.
   Note that, while the wave fronts are from an actual simulations, the perturbation shown in both panels is for illustration purposes only and their size is not to scale.
   \label{fig:comb_perturbs}}
\end{figure}

For each source location, we place a single flow perturbation pointing in $\theta$-direction (horizontal, southernly flow) on the same longitude but vary the flow perturbation depth and distance from the source, i.e., latitude, between simulations.
The flow perturbations are three-dimensional Gaussians with a FWHM of 1/4 of the local acoustic wavelength, except near the surface where the horizontal simulation resolution cannot resolve 1/4 wavelengh and a larger perturbation size of $1\fdg66$ FWHM is used instead.
However, the radial size of the flow perturbations is always 1/4 wavelength and is well resolved.
This is acceptable since near the surface waves propagate mostly radially.
The local wavelength $\lambda = c/\nu$ depends on the local sound speed $c$, which varies with depth, and the wave frequency $\nu$ for which we use a value of 3\,mHz\footnote{Later, we will make measurements not only at 3\,mHz but also at slightly larger frequencies.
For those larger frequencies, the ratio of perturbation size to wavelength will be slightly larger but still less than 1/2.}.
The peak velocity of the flow perturbations is set at 20\% of the local sound speed to maximize the amount of observable travel-time shift.
Such large flow speeds are acceptable since the governing equations in these simulations are linearized.
In order to cover depths down to 450\,Mm from the photosphere, well below the solar tachocline, and distances up to 45\degr, we need 784 different perturbations.
Their locations and sizes are shown in Figure~\ref{fig:perturbations}. Figure~\ref{fig:comb_perturbs} depicts one such perturbation and a snapshot of the wave front in order to better illustrate the geometry of the simulation setup.
Since each simulation contains two waves and two perturbations, one of each on each side of the simulated Sun, we compute 392 separate simulations, plus one extra simulation without flow perturbations as a reference for the travel-time measurements (see below).

\subsection{Frequency-Based Phase-Shift Measurements} \label{sec:measurements}
	
\begin{figure*}[t]
   \includegraphics[width=\linewidth]{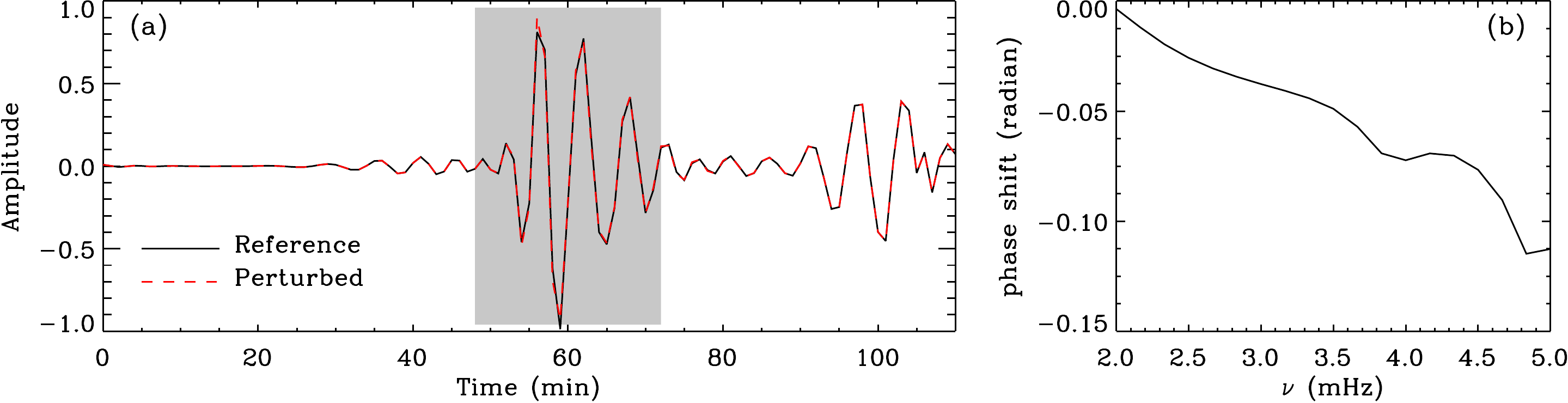}
   \caption{Process showing how $\delta\phi(\nu)$ is measured.
   (a) Full time series taken from a location in the reference simulation (solid line), and the corresponding series taken from the same location in a simulation with a flow perturbation (dashed red line).
   Only the first-skip sequences (marked by gray shading) are used for the phase-shift calculations.
   (b) Frequency-dependent phase shifts between the two curves in panel (a).
   \label{fig:phi_method}}
\end{figure*}

From the wave-field simulations we extract time series with 1\,minute cadence of the radial velocity at 300\,km above the photosphere, which approximately corresponds to the Doppler data observed by the Solar Dynamics Observatory/Helioseismic Magnetic Imager (SDO/HMI) \citep{2012SoPh..275..207S, 2012SoPh..275..229S}.
Specifically, from each wave simulation, around each wave-source location, we select a section of the surface that is $97\degr \times 97\degr$ in size ($\pm 48\fdg5$ in longitude around the source longitude, and from $-27\degr$ to $70\degr$ in latitude).
Since the waves in our simulations are launched by initial density perturbations and not by some finite-time forcing -- the source is a delta function -- we do not explicitly need to compute cross-correlation functions as is ordinarily done in time-distance helioseismology, but can use the signal at the receiver location directly as the signal in the phase-shift measurements.

In order to measure surface travel-time shifts $\dtau$, the time series at each location in the selected regions are compared with the time series taken at the same location from the reference simulation without perturbations (this corresponds to a point-to-point measurement geometry).
In observational analysis \citep[e.g.,][]{KosovichevDuvall1997}, helioseismic travel-times are usually measured by fitting the oscillatory signal with a Gabor-wavelet function which provides the phase travel time of the wave from the source location.
However, this method does not provide enough accuracy for the current work.
For more accurate measurements of the surface travel-time shifts, we take an alternative approach that is described as follows.
As shown in Figure~\ref{fig:phi_method}, we take a time sequence from the reference simulation and one sequence at the same location from a perturbed simulation, and select from them only the first-skip signal, i.e., the signal of the wave arriving the first time back at the surface.
This signal is approximately 25 minutes long in this simulation.
The two signals are then cross correlated in the Fourier domain yielding a frequency-dependent phase shift:
\begin{equation}
\delta\phi(\nu) = \arg \Big[ \mathcal{F}\big( \psi (\nu) \big) \cdot
\mathcal{F} \big( \psi_\mathrm{r} (\nu) \big)^\dagger \Big],
\label{eq5}
\end{equation}
where $\psi(t)$ and $\psi_\mathrm{r}(t)$ are the perturbed and reference wave signals, respectively.
The symbol $\mathcal{F}$ represents the Fourier transform, hence the oscillatory signals are converted from time $t$ domain to frequency $\nu$ domain.
To avoid introducing spurious signals into this short time sequence when performing Fourier transform, zero-padding with a length same to the first-skip signal is used, and signals near the boundaries of the sequence are slightly tapered. Note that this process does not enhance frequency resolution, which remains to be at about 0.33~mHz. Thus, when we speak of frequency 3.0~mHz, it actually means a frequency band centered at 3.0~mHz with a width of 0.33~mHz.
The $\dagger$ sign represents complex conjugation, and $\arg$ is the argument of the complex number, that is, the angle in the complex plane between the positive real axis and the line connecting the complex number with the origin.
The result is the phase shift between the two signals as a function of $\nu$, as shown in Figure~\ref{fig:phi_method}c.
The phase shift can be converted to a frequency-dependent travel-time shift using
\begin{equation}
\dtau(\nu) = \delta\phi (\nu) / 2\pi\nu.
\end{equation}
This method would fail for large travel-times shifts $|\dtau(\nu)| \ge 1/(2\nu) $ ($\sim 2.8$\, minutes for $3$\,mHz frequency) because of the $2\pi$-ambiguity of the complex argument, but this is not the case here since the travel-time shifts are only on the order of seconds.
Overall, this approach of deriving frequency-dependent travel-time shifts is similar to what \citet{ChenZhao2018} developed in their analysis, except that they used a long-time average function as their reference signal, but here we use the wave signal from a quiet-Sun simulation as our reference signal.

\begin{figure*}[t]
   \centering
   \includegraphics[width=0.65\linewidth,height=0.5\textheight,keepaspectratio]{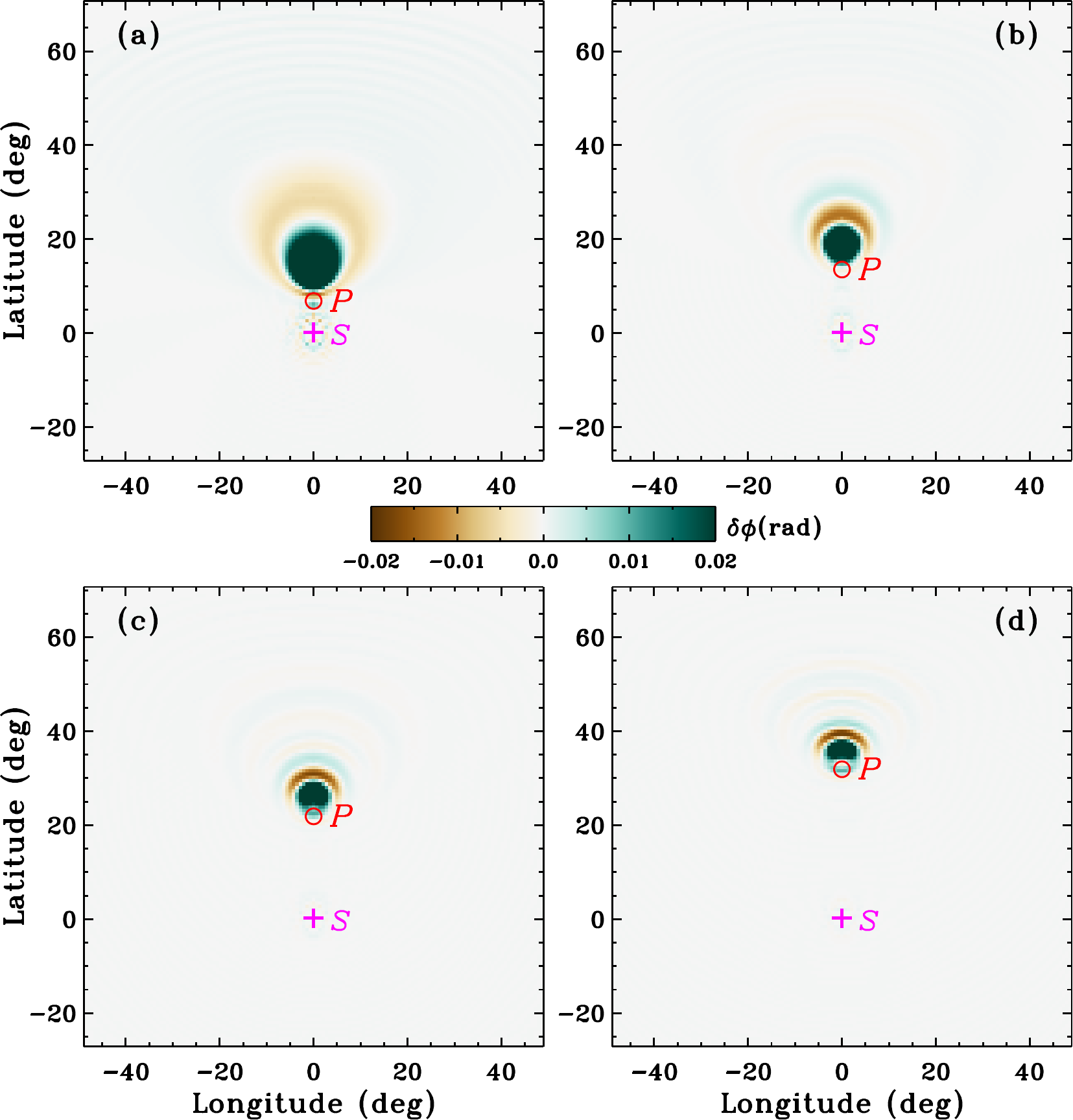}
   \caption{Select maps of phase shifts for $\nu = 3.0$\,mHz, for interior flow perturbations located at a depth of 54.5\,Mm and distances of (a) $6\fdg62$, (b) $13\fdg24$, (c) $21\fdg52$, and (d) $31\fdg46$ away from the wave source ${\cal S}$.
   For the cases shown, the perturbation size is $3.57$\,Mm FWHM in radius and $0\fdg7$ FWHM horizontally, and is indicated by circles marked $\cal P$.
   \label{fig:travel_time_map}}
\end{figure*}

Figure~\ref{fig:travel_time_map} shows examples of $\dtau$ maps for flow perturbations located at a depth of 54.5\,Mm and four select distances from the wave source.
It is found that despite the flow perturbations pointing southward, i.e., toward the wave source, $\dtau$ measured at the surface is not uniformly positive, as would be expected from ray-approximation theory, but shows zones of both positive and negative travel-time shifts, although positive values dominate in both magnitude and area.
Another fact worth pointing out is that the closer the perturbation is to the wave source, the wider the area where wave travel times are noticably shifted.
The area becomes quite small when the perturbation is far from the wave source, e.g., Figure~\ref{fig:travel_time_map}d; however, even in this case, the affected area is still a few times larger than the size of the perturbation.

\subsection{Solving for Sensitivity Kernels} \label{sec:kernels}
Each measurement point, for each simulation with a perturbation, corresponds to a convolution equation, like Equation~\ref{eqn:generickernel}, connecting the unknown kernel with the measured travel-time shifts.
However, equations for the same travel distance are coupled since the perturbations are not infinitesimally small, and therefore must be solved together.
This can be done numerically for a spatially discretized kernel.

\subsubsection{Discretized Kernel} \label{sec:kernels:discrete}
\begin{figure}[t]
  \centering
  \vspace{4mm}
  \includegraphics[viewport=24mm 24mm 80mm 82mm,clip,width=0.6\linewidth,height=0.25\textheight,keepaspectratio]{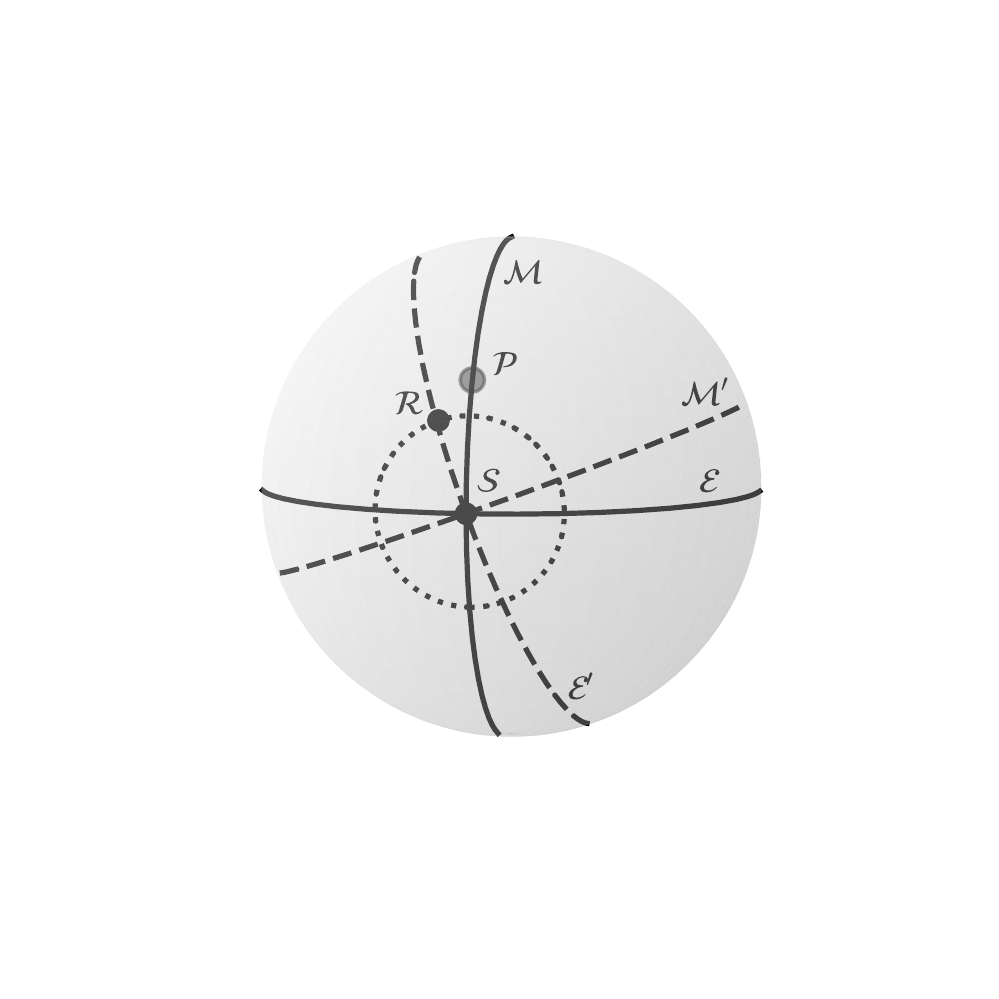}
  \vspace{2mm}
  \caption{Schematic illustration of the location of the wave source $\cal S$, receiver $\cal R$ and perturbation $\cal P$.
  Two types of coodinates are used here: Sun-fixed coordinates with $\cal E$ and $\cal M$ denoting the solar equator and the meridian on which the source is located, and kernel coordinates in which source and receiver lie on the equator, with ${\cal E}^\prime$ and ${\cal M}^\prime$ denoting equator and central meridian.
  Flow perturbations are always located on the same meridian $M$ as the wave source but lie at different latitudes and depths (see Section~\ref{sec:setup} and Figure~\ref{fig:perturbations}).
  For a given travel distance, any point along the dotted circle around the source is used as a receiver location.
  Although perturbation flows always point in latitudinal direction (in Sun-fixed coordinates), their  direction in kernel coordinates depends on receiver location and so does their distance from the kernel equator (transverse offset) and along the equator (longitudinal offset).
  \label{fig:geometry}}
\end{figure}

To facilitate solving for the sensitivity kernels, we define kernel coordinates ($r^\prime$, $\theta^\prime$, $\phi^\prime$) in which the wave source $\cal S$ and receiver $\cal R$ (i.e., the measurement point) lie on the equator of this new coordinate system.
Specifically, the wave source is located at $\phi^\prime=0$ and the receiver at $\phi^\prime=d$, with $d$ being the travel distance.
This coordinate system is in general different from the Sun-fixed coordinates ($r$, $\theta$, $\phi$) since we measure travel-time shifts on a large part of the solar surface.
Figure~\ref{fig:geometry} illustrates the two coodinate systems and how they relate to each other.
To avoid confusion between these coordinates, we call distances in kernel coordinates longitudinal offsets (distance from the source along the source-receiver line; in $\phi^\prime$ direction) and transverse offsets (distances perpendicular to the source-receiver line; in $\theta^\prime$ direction).
One important point is to realize that in the kernel coordinates the kernel is the same for all measurements with the same travel distance.
We can then define a discrete grid ($r^\prime_j$, $\theta^\prime_j$, $\phi^\prime_j$) in these coordinates with index $j$ running over all grid points.
The convolution equation (Equation~\ref{eqn:generickernel}) becomes a discrete sum:
\begin{equation}
   \delta\tau_i = \sum_j \left[ K^{\parallel}_j u^{\parallel}_{i,j} + K^{\perp}_j u^{\perp}_{i,j} \right] V_j
   \label{eqn:discretekernel}
\end{equation}
where $K^{\parallel}_j$ and $K^{\perp}_j$ are the unknown components of the kernel in the direction along the source-receiver line ($\phi^\prime$ direction) and perpendicular to it ($\theta^\prime$ directions), and $u^{\parallel}_{i,j}$ and $u^{\perp}_{i,j}$ are the two known components of the perturbation velocity, also discretized in kernel coordinates.
The $V_j$ are the volume elements of the discrete grid.
Notice that in the Sun-fixed coordinates we define the perturbations to have non-zero velocities only in latitudinal direction (Section~\ref{sec:setup}), but that in kernel coordinates they can have components in both horizontal directions depending on the receiver location.
The index $i$ in Equation~\ref{eqn:discretekernel} runs over all different measurements points for a given travel-distance $d$ and all different perturbations.
The equations for all $i$ combined represent a coupled set of linear equations -- a matrix equation:
\begin{equation}
   Ax = b,
\end{equation}
where the matrix is given by the perturbation velocities:
\begin{equation}
    A =
    \begin{bmatrix}
       u^{\parallel}_{1,1} & \dots & u^{\parallel}_{1,N_{\rm k}} & u^{\perp}_{1,1} & \dots & u^{\perp}_{1,N_{\rm k}} \\
       \vdots & \ddots & \vdots & \vdots & \ddots & \vdots \\
       u^{\parallel}_{N_{\rm m},1} & \dots & u^{\parallel}_{N_{\rm m},N_{\rm k}} & u^{\perp}_{N_{\rm m},1} & \dots & u^{\perp}_{N_{\rm m},N_{\rm k}}
    \end{bmatrix} \cdot V,
    \label{eqn:matrix}
\end{equation}
with the volume elements on the diagonal matrix
\begin{equation}
    V =
    \begin{bmatrix}
       V_1 &        &                 &     &        &                 \\
           & \ddots &                 &     &        &                 \\
           &        & V_{N_{\rm k}}   &     &        &                 \\
           &        &                 & V_1 &        &                 \\
           &        &                 &     & \ddots &                 \\
           &        &                 &     &        & V_{N_{\rm k}}
    \end{bmatrix},
\end{equation}
and where $b$ and $x$ are vectors of the measured travel-time shifts and the unknown kernel values:
\begin{equation}
    b = \begin{bmatrix} \delta\tau_1 \\ \vdots \\ \delta\tau_{N_{\rm m}} \end{bmatrix}, \,\,\,\,
    x = \begin{bmatrix} K^{\parallel}_1 \\ \vdots \\ K^{\parallel}_{N_{\rm k}} \\ K^{\perp}_1 \\ \vdots \\ K^{\perp}_{N_{\rm k}} \end{bmatrix}.
\end{equation}

We select a kernel domain that is large enough to encompass the entire region where the kernel may have significant sensitivity to flows.
Specifically, we use longitudinal offsets $\phi^\prime \in [-45\degr, d + 45\degr]$, transverse offsets $\theta^\prime - 90\degr \in [-45\degr, 45\degr]$, and depths $R_\odot - r^\prime \in [0, 450]$\,Mm.
Recall that the prescribed perturbations (section~\ref{sec:setup}) are up to 45\degr away from the source, and so only travel distances $d$ a bit less than this distance can be considered here.
This is also the reason why we do not consider larger transverse offsets.
Similarly for the longitudinal offsets, note that the perturbations only cover offsets $\phi^\prime \in [-45\degr, 45\degr]$ and the results beyond ($\phi^\prime \in [45\degr, d + 45\degr]$) are obtained by symmetrization, i.e., they originate from offsets $\phi^\prime \in [-45\degr, d - 45\degr]$.
Longitudinal and transverse offsets are discretized with an equidistant grid with a spacing of $\Delta\sim0\fdg7$, the same resolution as the wave-propagation simulations, or slightly better so that the range of longitudinal and transverse offset is an exact multiple of the resolution. Also, we make sure that there is a grid point at zero transverse offset and at the midpoint of the travel distance in (for easy symmetrization of the kernel later).
In the radial direction, we use 54 points on a stretched grid such that the grid spacing is approximately proportional to the local sound speed.

\subsubsection{Selection of Measurement Points} \label{sec:kernels:selection}

\begin{figure*}[t]
   \centering
   \includegraphics[width=\linewidth]{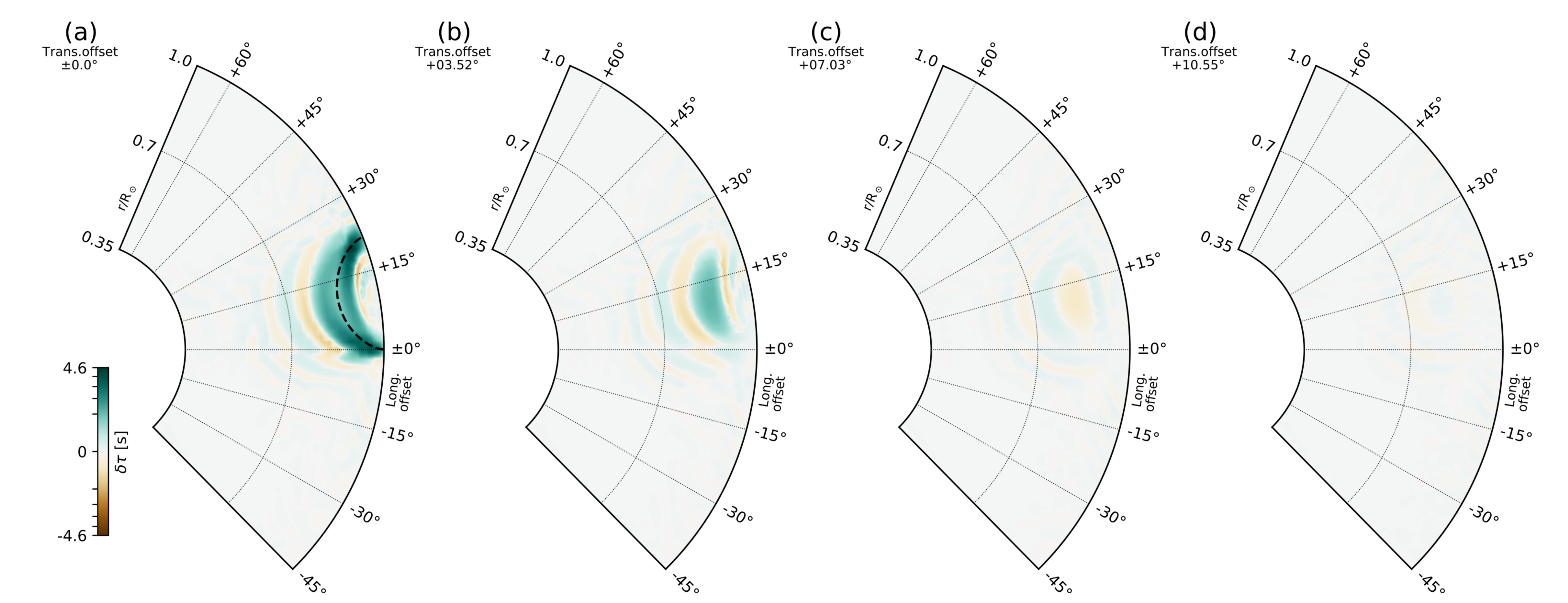}
   \caption{Travel-time shifts in kernel coordinates obtained by remapping and combining surface measurement for all perturbations and all locations with a distance of $d=21\fdg6$ from the source, at a frequency of $3$\,mHz.
   The panels (a)--(d) show cross-sections at select transverse offsets as indicated in each panel.
   A dashed line in panel (a) represents the ray path connecting the wave source and receiver.
   \label{fig:measurements_in_kernel_coordinates}}
\end{figure*}

For a given travel distance $d$, we select measurement points along the circle of radius $d$ around the source location (as illustrated by the dotted circle in Figure~\ref{fig:geometry}).
We choose points on the circle in steps of the simulation resolution ($\Delta\sim0\fdg7$) and use third-order spline interpolation of the gridded measurements (Section~\ref{sec:measurements}).
For each point, we then compute the location and direction of the flow perturbation in kernel coordinates ($\theta^\prime, \phi^\prime$), and populate the discrete grid with the flow perturbation $u^{\parallel}_{i,j}$ and $u^{\perp}_{i,j}$.
Since a circle of radius $d$ on the unit sphere has a circumference of $2\pi\sin(d)$, there are approximately $2\pi\sin(d)/\Delta$ measurements points along the circle.
In practice, however, some of these points can not be used because they may lie outside the measurement region, or the flow perturbation may lie outside the kernel grid (remember that the kernel grid moves with the measurement point; measurement point, or receiver $\cal R$, and source $\cal S$ always lie on the equator in kernel coordinates).

Before proceeding to solving the coupled set of equations, let us have a look at the measurements from Section~\ref{sec:measurements} plotted in kernel coordinates.
Specifically, Figure~\ref{fig:measurements_in_kernel_coordinates} shows the travel-time shifts of all measurement points with a travel distance of $d=21\fdg6$ and all different perturbations, and plots them at their perturbation location in kernel coordinates.
For infinetesimal perturbations, this would be identical to the kernel itself except for some constant factors.
By plotting the measurements this way, it is apparent that waves have non-zero sensitivity to perturbations outside the ray path, which is indicated by the dashed line in Figure~\ref{fig:measurements_in_kernel_coordinates}a, as expected, and in contrast to ray-approximation kernels.
Still, $\dtau$ drops fairly quickly with distance away from the source-receiver plane.
Overall, the travel-time shifts could be described as a ``banana''-shaped region, positive in the central region, with layers of alternating signs around it.
The ray path connecting the wave source and receiver passes through the the middle of this banana-shaped region.
Of note is that the travel-time shifts caused by perturbations near the wave source and near the wave receiver are different, and therefore travel-time shifts are not symmetric with respect to the travel direction.
Ultimately, we will symmetrize the kernels derived from these measurements because for observations we typically measure travel-time differences between oppositely traveling waves.

\subsubsection{Regularization and Solver}  \label{sec:kernels:regularization_and_solver}
Since the linear system is strongly underdetermined, in order to compute physically-reasonable, smooth solutions, we introduce a regularization term with parameter $\lambda$.
In matrix form, this can be accomplished by replacing $A$ and $b$ with:
\begin{equation}
    A \to \tilde{A} =
    \begin{bmatrix} A \\ \lambda\sqrt{V} \\ \end{bmatrix}, \,\,\,\,
    b \to \tilde{b} = \begin{bmatrix} b \\ 0 \\ \vdots \\ 0 \end{bmatrix}.
\end{equation}
When solving $\tilde{A} x = \tilde{b}$ in a least-squares sense, i.e., minimizing $||\tilde{A}x-\tilde{b}||^2$, the regularization term in effect reduces the volume integral of the squared kernel (for sufficiently large $\lambda$).
This makes the solution smoother.
In order to balance smoothness and accuracy, we iteratively change $\lambda$ starting with some initial value, compute a solution and its residual ${\cal R}=\sqrt{||A x-b||^2}$, and then increase or decrease $\lambda$ until we reach a value of $\lambda$ that minimizes the residual.

The least-squares solutions are computed using the iterative solver LSQR \citep{PaigeSaunders1982a} available from the Python library SciPy \citep{2020SciPy-NMeth}.
LSQR can take advantage of sparse matrices.
Realizing that very small values of the perturbation velocities do not contribute significantly to the travel-time shifts, we can increase the sparesness of the problem by setting $u^{\parallel}_{i,j}$, $u^{\perp}_{i,j}$ to zero wherever their value is less than, say, $10^{-3}$ times their maximum value.
Typically, only $0.01$\,\% of the matrix elements are then non-zero.
This dramatically reduces the number of non-zero values in the matrix, reducing both storage requirements and speeding up the calculations without significantly affecting the results.
Without this ``trick" such calculations would not be possible even on the largest of supercomputers.
For instance, the kernels for $21\fdg6$ and $36\degr$ travel distance shown in the results section require solving linear systems with sizes of $2,390,444 \times 2,243,052$ and $2,756,892 \times 2,521,692$,\footnote{The second dimension is the number of unknowns, i.e. the number of discretization point of the kernel (times two since we have two components -- longitudinal and transverse). The first dimension is the number of equations, this is the number of measurements (~$2\pi\sin(d)/\Delta$, see section~\ref{sec:kernels:selection}, times the number of perturbations) plus the numbers of unknowns due to the regularization term.} but only require a few GB of storage and computing a single solution on a 20-core machine takes about 7 and 12\,min, respectively.

\subsubsection{Symmetrization}
As a last step, we symmetrize the kernels about the midpoint of the travel distance, $\phi^\prime = d/2$.
This is done because for observations we typically measure travel-time differences between waves going along opposite directions, i.e., $\cal S$-to-$\cal R$ vs $\cal R$-to-$\cal S$ in Figure~\ref{fig:geometry}, while the travel-time shifts measured on the simulation are just in one travel direction relative to the reference state without perturbations.
By symmetrizing the kernel we effectively get the sensitivity of travel-time shifts between oppositely traveling waves.
It also removes differences between the reference state in the simulation and the mean Sun assuming that the differences are small.

\section{Results} \label{sec:results}

\begin{figure*}[t]
   \centering
   \includegraphics[width=\linewidth]{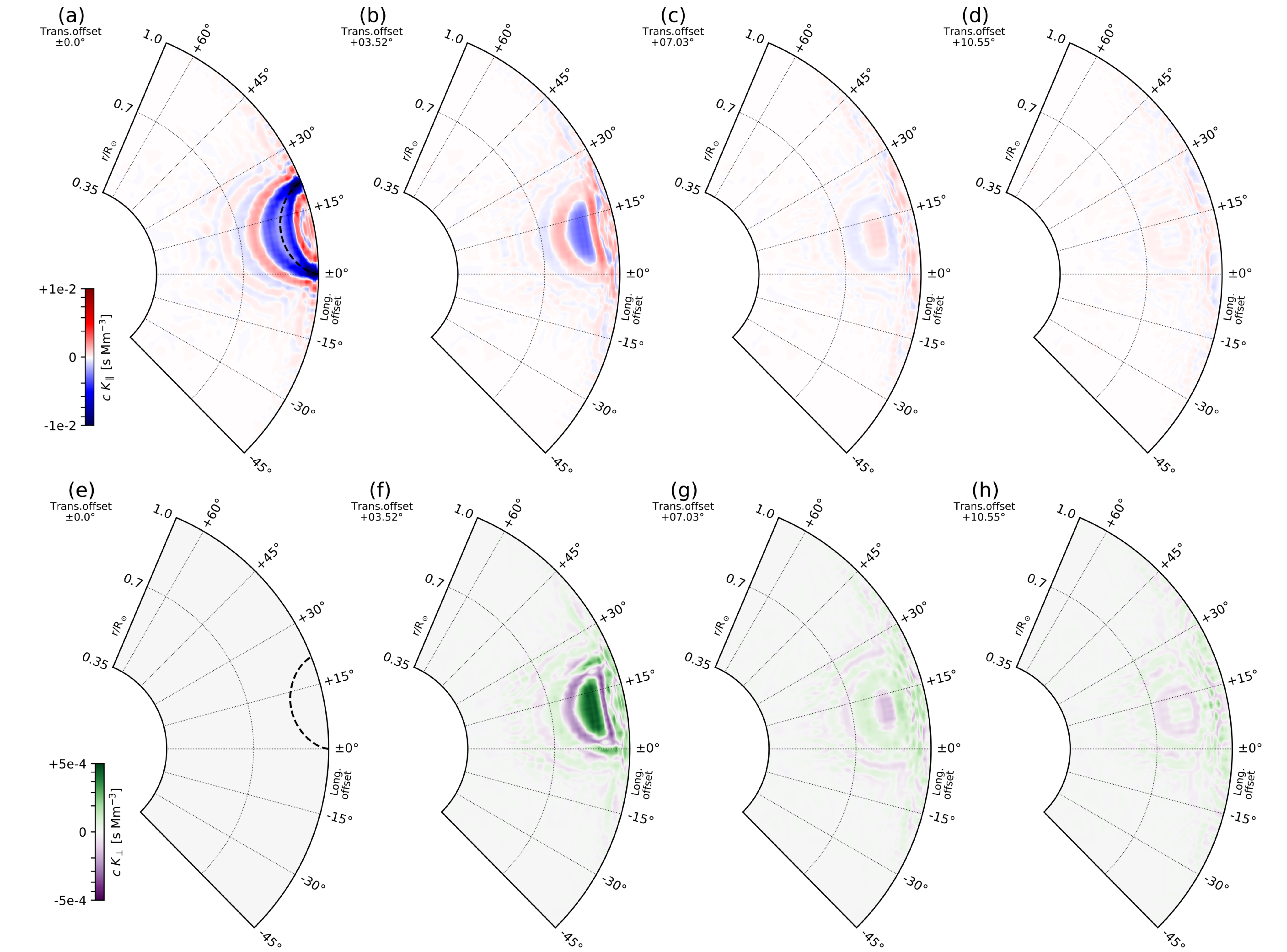}
   \caption{Sensitivity kernels (multiplied with sound speed $c$) for longitudinal flows (upper row) and
   transverse flows (lower row) for a measurement distance of $21\fdg6$ and a frequency of $\nu=3$\,mHz.
   The different panels in each row show vertical slices of the kernel at transverse offsets from the central plane of $0\fdg0$, $3\fdg52$, $7\fdg03$, and $10\fdg55$ (from left to right), respectively.
   The dashed lines in panels (a) and (e) represent the ray path connecting the two surface locations between which the travel-time shifts are measured.
   The color scale used within each row is the same but note that they are streched in order to make small-amplitude values more visible and oversaturates the near-surface kernels.
   \label{fig:kernel1}}
\end{figure*}

\begin{figure*}[p]
   \centering
   \includegraphics[width=\linewidth]{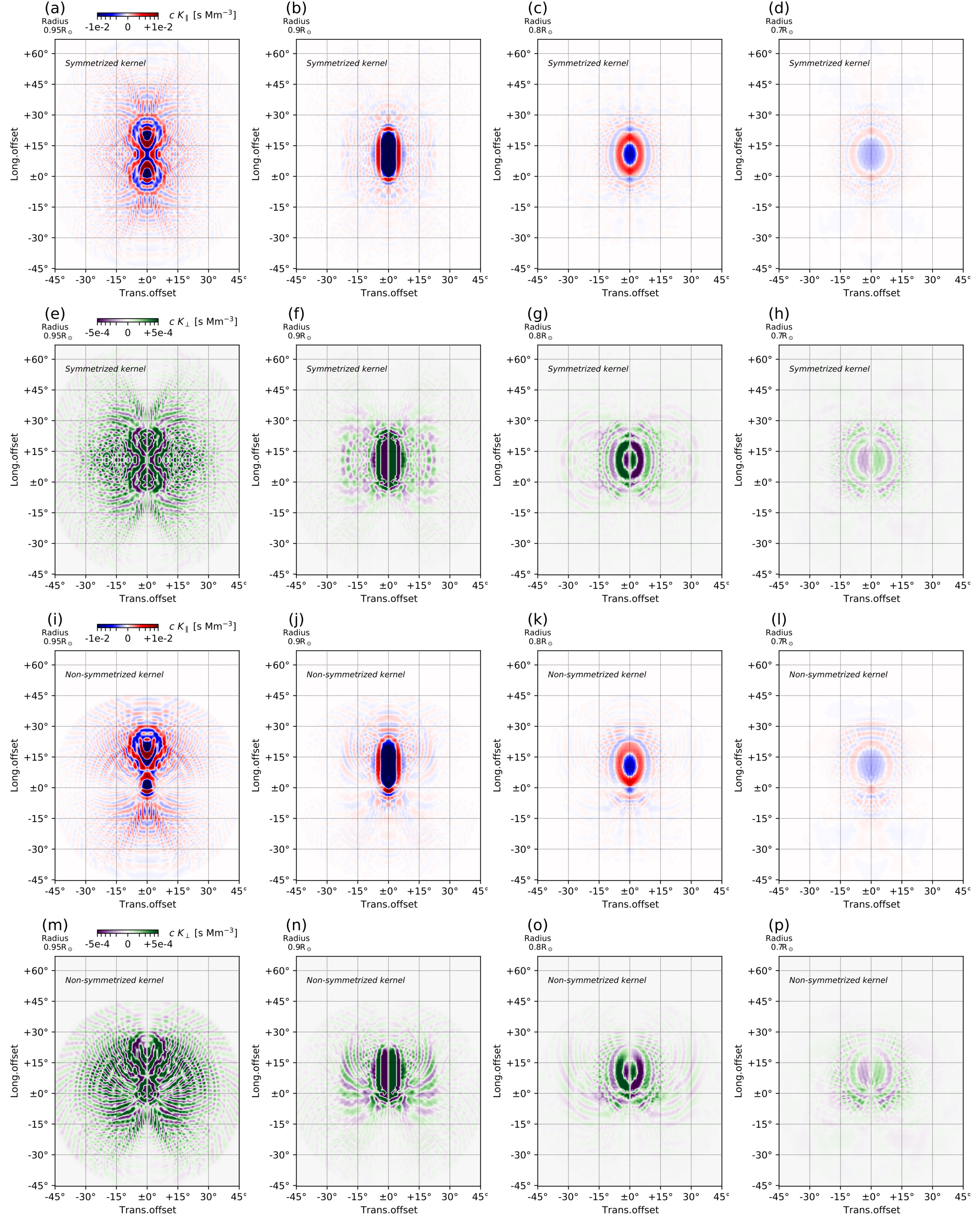}
   \caption{Horizontal slices of the sensitivity kernels for longitudinal flows (panels a--d and i--l) and transverse flows (panels e--h and m--p) for a measurement distance of $d=21\fdg6$ and a frequency of $\nu=3$\,mHz. The first two rows (panels a--h) show the symmetrized kernel while the last two rows (panels i--p) show the kernel before symmetrization. Again, the kernels have been multiplied with sound speed $c$ to make small values more visible.
   \label{fig:h_kernel1}}
\end{figure*}

\begin{figure*}[p]
   \centering
   \includegraphics[width=\linewidth]{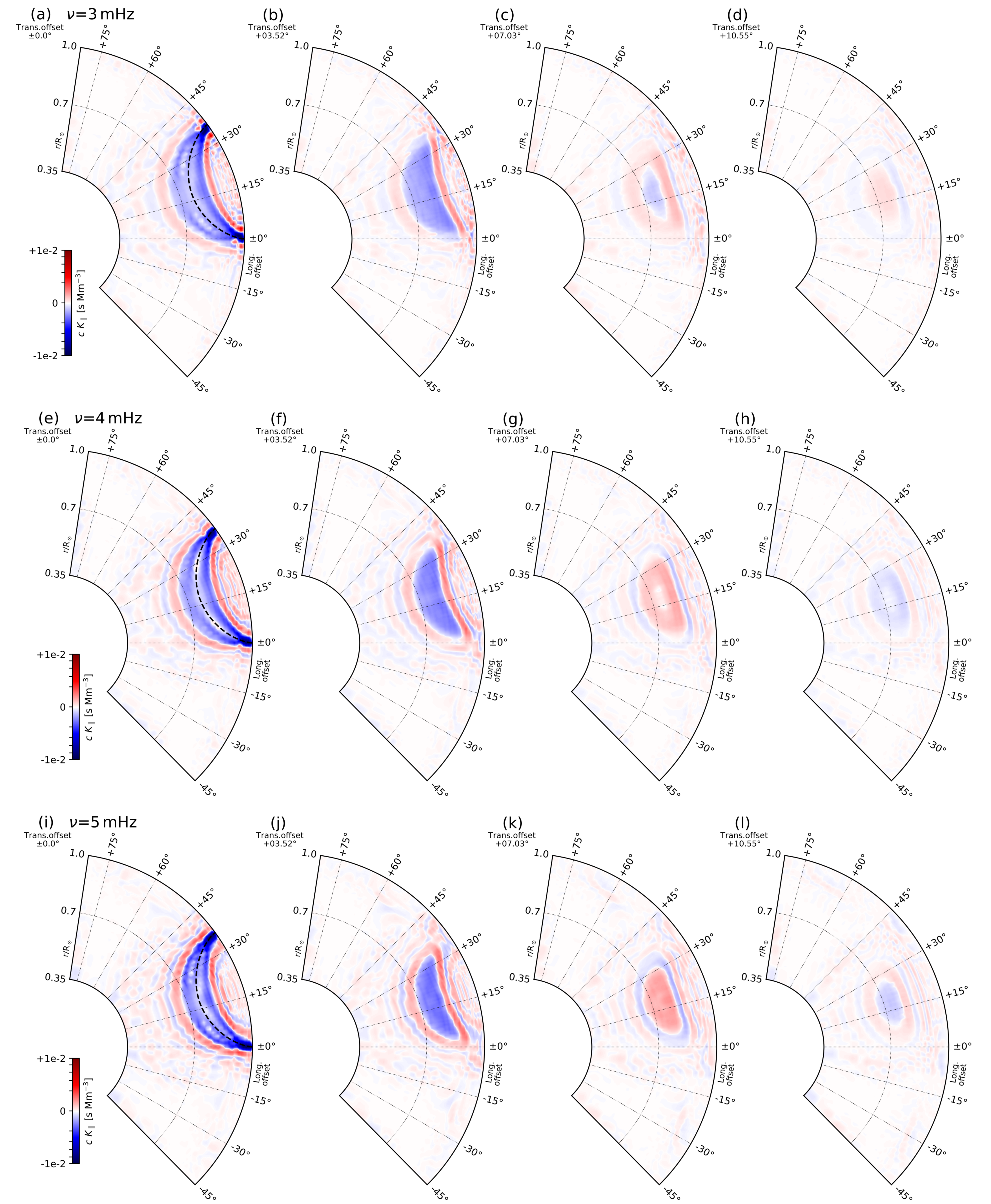}
   \caption{Longitudinal component of the flow sensitivity kernel (multiplied with sound speed $c$) for a measurement distance of $36\degr$ at transverse offsets of $0\fdg0$, $3\fdg52$, $7\fdg03$, and $10\fdg55$ (from left to right) and frequencies of $3$, $4$ and $5$\,mHz (from top to bottom), respectively.
   \label{fig:lon_kernel_234}}
\end{figure*}

\begin{figure*}[p]
   \centering
   \includegraphics[width=\linewidth]{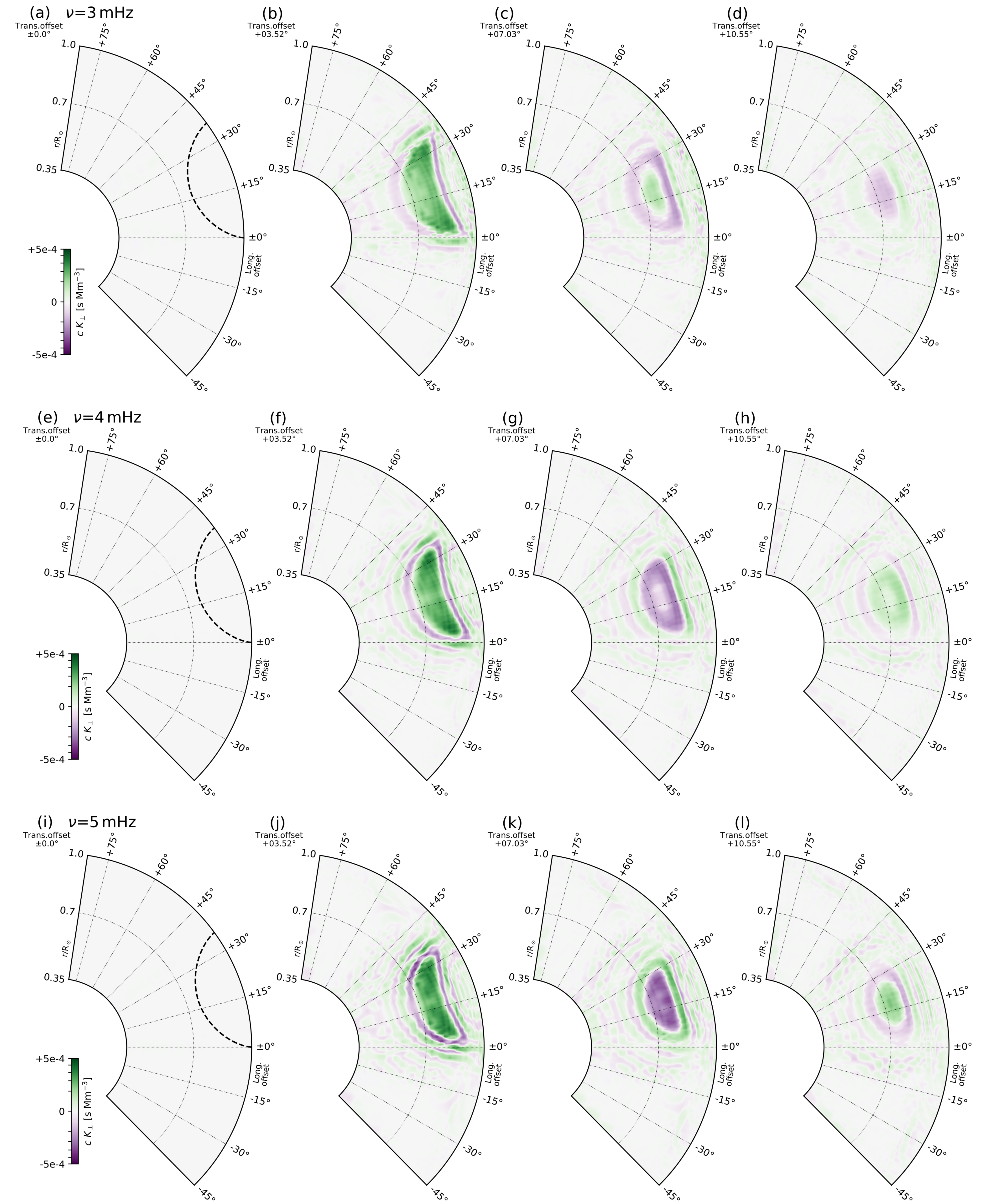}
   \caption{Transverse component of the flow sensitivity kernel (multiplied with sound speed $c$) for a measurement distance of $36\degr$ at transverse offsets of $0\fdg0$, $3\fdg52$, $7\fdg03$, and $10\fdg55$ (from left to right) and frequencies of $3$, $4$ and $5$\,mHz (from top to bottom), respectively.
   \label{fig:tra_kernel_234}}

\end{figure*}

Figures~\ref{fig:kernel1} and \ref{fig:h_kernel1} show, respectively, cross-sectional and horizontal views of sensitivity kernels for both longitudinal and transverse flows, for a measurement distance of $21\fdg6$ and a frequency of 3\,mHz at select offsets or depths.
Figures~\ref{fig:lon_kernel_234} and \ref{fig:tra_kernel_234} show cross-sectional views of sensitivity kernels for longitudinal flows and transverse flows, respectively, for a measurement distance of $36\degr$ and frequencies of 3, 4, and 5 mHz at select offsets.

Consistent with previous works on Born kernels, it is found that travel-time shifts measured on the surface are most sensitive to the longitudinal flows located on the same vertical plane as the ray path.
With increasing offset distance from the central plane, the sensitivity rapidly drops and starts to alternates in sign.
Similarily, the sensitivity becomes weaker and alternates in sign for flows above and below a banana-shaped region that is roughly centered at the ray path.
Although the amplitude decreases with depth, sensitivity reaches quite far below the ray path.
Note that the longitudinal kernel is symmetric with respect to the central plane (see Figure~\ref{fig:h_kernel1}a--d).
This is true even before explicit symmetrization (see Figure~\ref{fig:h_kernel1}i--l).
However, the non-symmetrized kernel is different near the source and receiver as has been found by previous works~\citep[e.g.,][]{GizonBirch2002, Mandaletal2017}.
Figure~\ref{fig:h_kernel1}i--l show that this is in particular true for the near-surface kernel while the asymmetry significantly diminishes deeper in the interior.

It has been known that travel-time shifts have non-zero sensitivity to transverse flows \citep[e.g.,][]{BirchGizon2007}.
Figures~\ref{fig:kernel1}e--h, ~\ref{fig:h_kernel1}e--h and m--p, and \ref{fig:tra_kernel_234} show that the sensitivity is about one order of magnitude weaker than for longitudinal flows.
The sensitivity kernels for transverse flows are antisymmetric with respect to the central plane (see Figure~\ref{fig:h_kernel1}e--h and m--p), which is easily understood geometrically.
The sensitivity is zero on the central plane; it first increases with increasing transverse offset for some distance, and then drops again for larger offsets, alternating in sign similarly to the sensitivity kernels for longitudinal flows.

It is also noteworthy, albeit unsurprisingly, that the sensitivity kernels change with frequency (see Figures~\ref{fig:lon_kernel_234} and \ref{fig:tra_kernel_234}).
With increasing frequency, the wavelength of helioseismic waves decreases, and similarily the thickness of the central banana shape and the spacing between alternating signs in the sensitivity decreases.
For the same mass flow, travel-time shifts measured at a higher frequency are slightly larger than those measured at a lower frequency.

In comparison to previous works \citep[e.g.,][]{Boeningetal2016, Mandaletal2017}, our kernels show less high-amplitude small-scale ``ringing'' near the solar surface.
We suspect that this may be due to the limited spatial resolution of our numerical simulations.
In the interior near the ray path our kernels are of similar magnitude as those shown by \citet[][their figure~5]{Boeningetal2016} which however seem to disagree with \citet[][their figure~7]{Mandaletal2017}.
Although, we suspect that \citet{Mandaletal2017} may have missquoted their units as s\,cm$^{-3}$ while they were in fact s\,Mm$^{-3}$ as in our figures.
In this case, their results would be consistent with our kernels.

\section{Validation} \label{sec:validation}

\begin{figure*}[t]
   \centering
   \includegraphics[width=\linewidth]{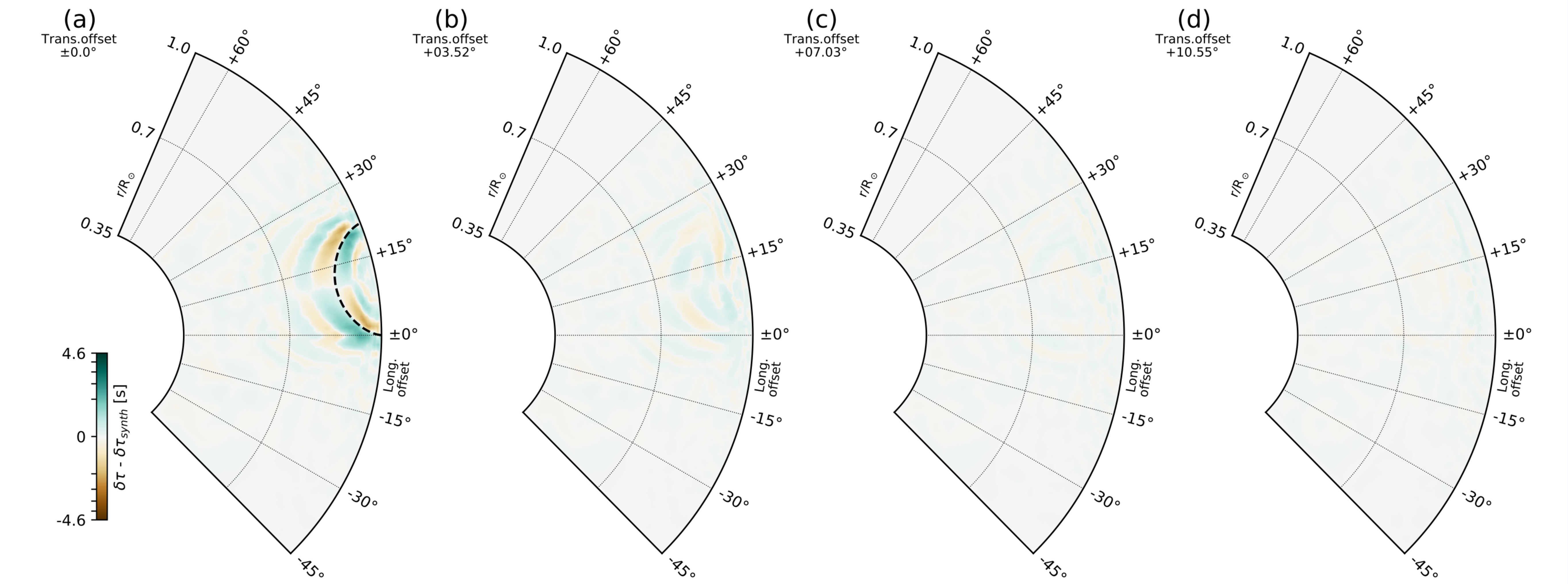}
   \caption{Difference between measurements, $\delta\tau$, and synthetic measurements, $\delta\tau_{synth}$, computed by convolution of the flow perturbations with the new (non-symmetrized) sensitivity kernels for a travel distance of $d=21\fdg6$ and a frequency of $3$\,mHz. The color scale is the same used in figure~\ref{fig:measurements_in_kernel_coordinates}.
   As before, panels (a)--(d) show cross-sections at select transverse offsets as indicated in each panel.
   A dashed line in panel (a) represents the ray path connecting the wave source and receiver.
   \label{fig:synthetic_measurements_in_kernel_coordinates}}
\end{figure*}

In order to demonstrate that these results are in fact accurate, we performed checks for self-consistency and compared travel times predicted from the present kernels with ray-approximation calculations.

A straighforward check for self-consistency can be done by using the new kernels to compute the expected travel times for the flow perturbations used in the simulations.
These ``synthetic measurements'' can then compared to the actual travel-time measurements discussed in section~\ref{sec:measurements}.
For obvious reasons, we use the non-symmetrized kernels for these calculations.
Figure~\ref{fig:synthetic_measurements_in_kernel_coordinates} gives an example of these synthetic measurements for one travel distance and frequency showing the difference between synthetic measurements and the corresponding measurements from figure~\ref{fig:measurements_in_kernel_coordinates}.
Although the maximum difference is more than 1 sec, the differences are mostly antisymmetric with respect to the midpoint of the travel distance.
Therefore, when measuring round-trip travel-time differences, as is usually done in practice, the error is much smaller.

\begin{figure*}[t]
   \centering
   \hspace{-0.020\linewidth}

   \includegraphics[viewport=0mm 0mm 71mm 125mm,clip,height=0.45\linewidth]{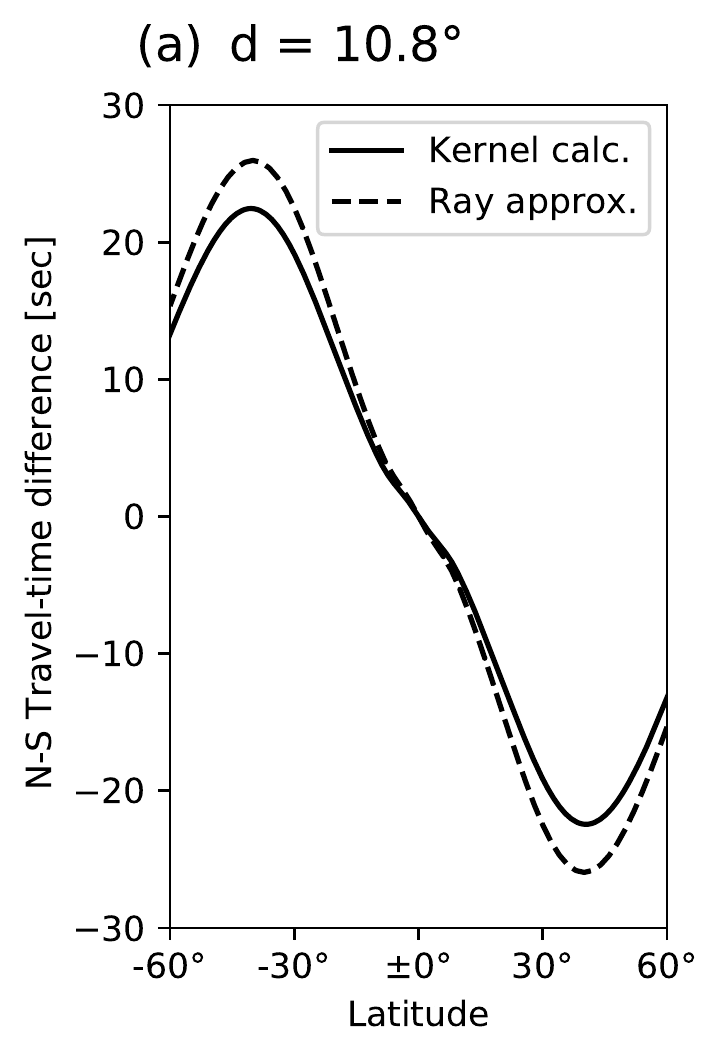}
   \includegraphics[viewport=8mm 0mm 71mm 125mm,clip,height=0.45\linewidth]{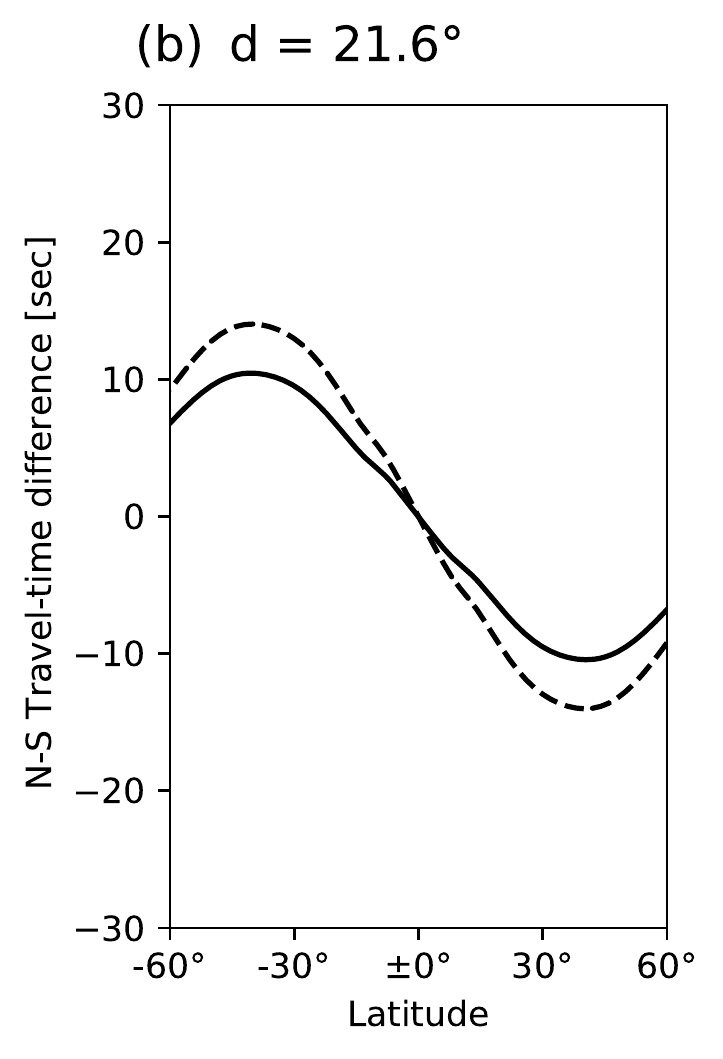}
   \includegraphics[viewport=8mm 0mm 71mm 125mm,clip,height=0.45\linewidth]{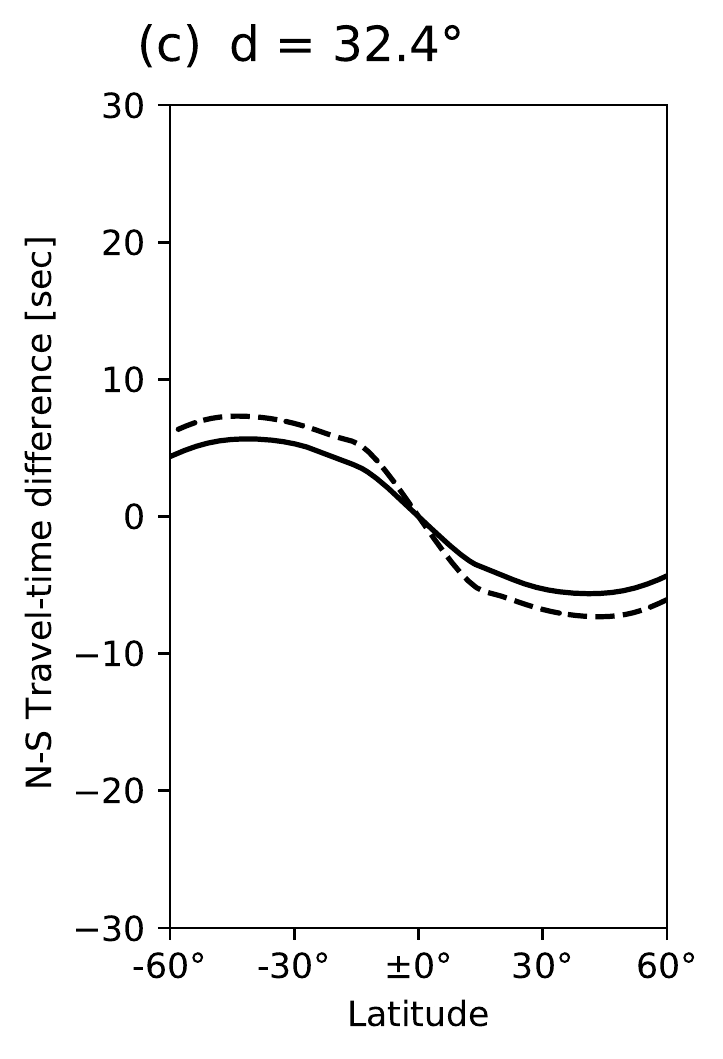}
   \includegraphics[viewport=8mm 0mm 71mm 125mm,clip,height=0.45\linewidth]{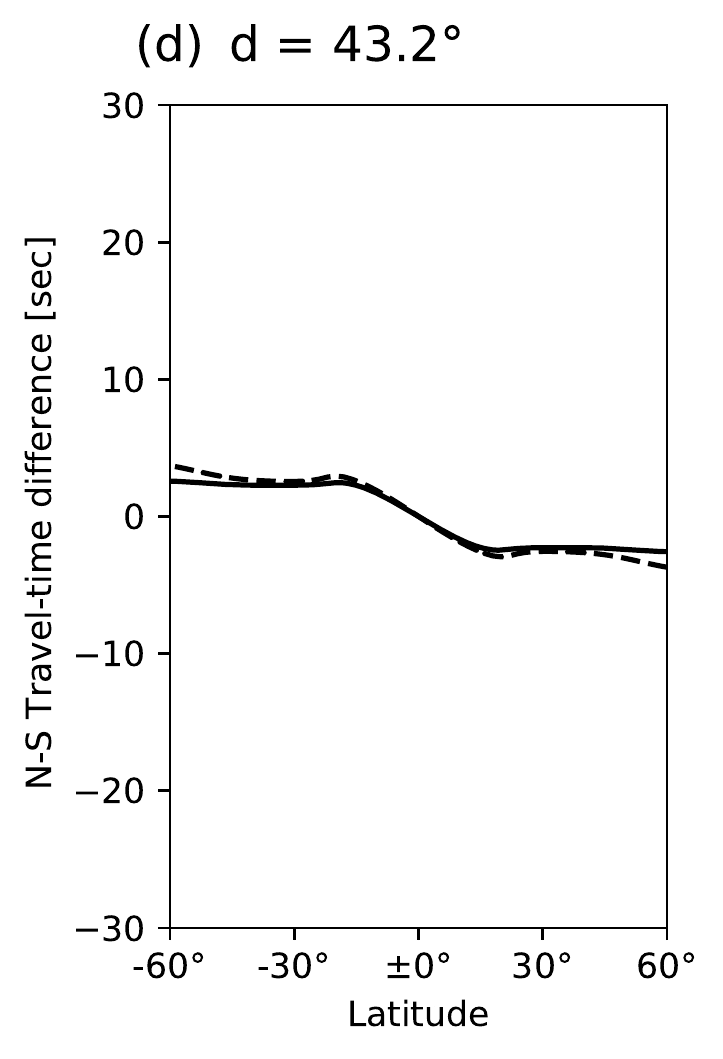}

   \caption{Forward calculations of expected travel-times from the meridional flow of \citet{Hartlepetal2013} comparing results from the present sensitivity kernels with ray approximation calculations, both for a frequency of 3\,mHz.
   \label{fig:forward_problem}}
\end{figure*}

Another test of the new kernels in shown in Figure~\ref{fig:forward_problem}.
Using our symmetrized kernels (after appropriate rotation and azimuthal averaging) we compute the expected travel-times shifts caused by a meridional flow \citep[the model from][]{Hartlepetal2013}, and compare them with ray-approximation calculations.
We expect them to be somewhat different but of similar magnitude, which is what we found.
Given that the realistic kernels have finite extent and alternating signs, it is maybe not surprising that the travel-times are smaller than those predicted by ray approximation (as seen in Figure~\ref{fig:forward_problem}).

\section{Discussion} \label{sec:discussion}

Through measuring travel-time shifts in simulated wave-fields with pre-defined localized flow perturbations in the solar interior, we have been able to derive three-dimensional helioseismic sensitivity kernels for the Sun's horizontal internal flows.
These kernels have both a longitudinal component (for flows parallel to the wave travel direction) and a transverse component (for flows perpendicular to the wave travel direction) although the former is about one order of magnitude stronger than the latter.
These new flow sensitivity kernels are qualitatively similar to the kernels obtained using Born approximation by other authors \citep[e.g.,][]{Boeningetal2016, Gizonetal2017}, at least in the general shape of the kernels, demonstrating that both the Born-approximation method and our numerical-simulation method capture the basic physics of the interaction between the Sun's low- and medium-$\ell$ waves and its large-scale internal flows.
Our method provides a complementary approach to calculating sensitivity kernels, helping to increase the confidence in Born approximation kernels and offering alternative kernels for assessing their accuracy.
Clearly, quantitative comparisions of kernels from the different methods are a worthwile future avenue of work although care has to be taken what filters (frequeny or phase-speed) were considered as they crucially affect the kernels.
Another topic for future studies is the difference between single-source kernels (as derived here) and distributed-source kernels.
\citet{GizonBirch2002} have shown that there is a difference, but how important that is in practice remains to be seen.

The idea of deriving sensitivity kernels through measuring travel-time shifts by localized perturbations was previously explored by a few authors.
For example, \citet{Duvalletal2006} inferred sensitivity kernels of magnetic field for $f$-mode waves through measuring travel-time shifts caused by the $\delta$-function-like surface magnetic elements.
\citet{Hanasogeetal2007} attempted to derive sensitivity kernels for variations in the speed of sound through a numerical simulation with multiple pre-defined sound-speed perturbations, and then disentangled the effect of one localized perturbation from multiple perturbations.
Our method is different in that we carry out multiple simulations with only two localized perturbations per simulation placed far apart so that they are essentially independent, and this greatly simplifies the computation of kernels.
In this article, we only present results for the two horizontal components of flows, but our procedure is easily applied to radial flows or other internal properties on the Sun such as sound-speed perturbations or magnetic fields.
It is also worth mentioning that by far most of the computational cost of our procedure is in calculating the many wave-propagation simulations through localized perturbations, while the travel-time measurements and the calculation of the kernels is negligible in comparison.
Therefore, it is easy and computational inexpensive to compute kernels for other measurements procedures or, e.g., different frequency filters.

Sensitivity kernels for different frequencies are naturally obtained from our method using the same set of simulations.
All previous inversions for the Sun's meridional circulation have used broad-band travel-time fittings, which primarily come from the dominant frequency, although travel-time shifts of the waves vary with frequency.
Introducing frequency-dependent phase-shift measurements \citep[e.g.,][]{ChenZhao2018} and coupling these measurements with the frequency-dependent sensitivity kernels, as calculated in this work, will increase the number of indepenent measurements for a given measurement distance and should improve the reliability and robustness of the inversion results.
We plan to do this next for the inferrences of the Sun's meridional circulation.

\acknowledgments
This work was partly supported by NASA's Heliophysics Supporting Research program under grants NNX15AL64G and 80NSSC19K0857.
The authors thank the International Space Science Institute (ISSI) at Bern, Switzerland for hosting a workshop where part of this work was presented, discussed, and improved based upon comments received.
We are also grateful to the referee for their thorough review and helpful suggestions for improving the paper.

\bibliographystyle{aasjournal}
\bibliography{heliophysics_bibliography}

\end{document}